\documentclass[10pt,twocolumn]{IEEEtran}

\usepackage{color}
\usepackage{amsfonts}
\usepackage{subfigure}
\usepackage{amsmath}
\usepackage{amssymb}
\usepackage{epsfig}
\usepackage{mathrsfs}
\usepackage{graphicx}
\usepackage{mathrsfs}
\usepackage{dsfont}
\usepackage{algorithm}
\usepackage{caption}
\usepackage{algpseudocode}

\DeclareMathAlphabet{\baz}{OML}{cmm}{b}{i}

\def\bA{\mbox{\boldmath $A$}}
\def\bB{\mbox{\boldmath $B$}}
\def\bC{\mbox{\boldmath $C$}}
\def\bD{\mbox{\boldmath $D$}}

\def\bF{\mbox{\boldmath $F$}}

\def\bI{\mbox{\boldmath $I$}}

\def\bL{\mbox{\boldmath $L$}}

\def\bP{\mbox{\boldmath $P$}}

\def\bW{\mbox{\boldmath $W$}}

\def\b0{\mbox{\boldmath $0$}}

\def\bb{\mbox{\boldmath $b$}}

\def\bh{\mbox{\boldmath $h$}}
\def\bh{\mbox{\boldmath $h$}}

\def\bu{\mbox{\boldmath $u$}}
\def\bw{\mbox{\boldmath $w$}}
\def\bx{\baz{x}}

\def\by{\mbox{\boldmath $y$}}
\def\bz{\mbox{\boldmath $z$}}

\newenvironment{proof}[1][Proof]{\noindent \textbf{#1.} }{\qedsymbol}
\newcommand{\qedsymbol}{\hspace{\fill}\rule{1.5ex}{1.5ex}}

\title{Distributed Estimation and Control of Algebraic Connectivity over Random Graphs}

\author{Paolo~Di Lorenzo,~\IEEEmembership{Member,~IEEE}, and Sergio~Barbarossa,~\IEEEmembership{Fellow,~IEEE}\\
\thanks{This work has been supported by TROPIC Project, Nr. 318784. The authors are with the Department of Information, Electronics, and Telecommunications, ``Sapienza'' University of Rome, Via Eudossiana 18, 00184 Rome, Italy. e-mail: {\tt \{dilorenzo,sergio\}@infocom.uniroma1.it}. Part of this work was presented at the International Conference on Acoustics, Speech, and Signal Processing (ICASSP), Vancouver, May 2013 \cite{DiLo_Barb4}.}}

\begin{document}

\maketitle

\begin{abstract}
In this paper we propose a distributed algorithm for the estimation and control of the connectivity of ad-hoc networks in the presence of a random topology. First, given a generic random graph, we introduce a novel stochastic power iteration method that allows each node to estimate and track the algebraic connectivity of the underlying expected graph. Using results from stochastic approximation theory, we prove that the proposed method converges almost surely (a.s.) to the desired value of connectivity even in the presence of imperfect communication scenarios. The estimation strategy is then used as a basic tool to adapt the power transmitted by each node of a wireless network, in order to maximize the network connectivity in the presence of realistic Medium Access Control (MAC) protocols or simply to drive the connectivity toward a desired target value. Numerical results corroborate our theoretical findings, thus illustrating the main features of the algorithm and its robustness to fluctuations of the network graph due to the presence of random link failures.
\end{abstract}


\section{Introduction}

Ad-hoc wireless networks are composed of a set of nodes that exchange data with each other through wireless links without relying on any pre-existing infrastructure \cite{Barb-Sard-Dilo}. Typically, the network topology follows a nearest neighbor criterion to allow for low-power communications.    The diffusion of information through a network presumes connectivity of the network. Furthermore, many distributed algorithms running over a graph, such as consensus, diffusion, or swarming algorithms, have a convergence time strictly dependent on the graph connectivity \cite{Xiao-Boyd}-\cite{DiLo_Barb3}. For example, highly connected networks generally have significantly faster convergence thanks to a more efficient in-network information diffusion.  In many practical examples, connectivity can only be assumed to hold in probability because the links among the nodes may be on or off depending on channel conditions. In most applications, channel variability may depend on several factors, such as mobility of the nodes, as in vehicular networks, channel fading due to propagation over multipath channels, or packet collisions due to random MAC strategies working on a collision avoidance regime. It is then of interest to look at distributed mechanisms to estimate and control the network connectivity in the presence of realistic channel models.

Spectral graph theory \cite{Chung} has been demonstrated to be a very powerful tool for topology inference. The eigenvalues and/or eigenvectors of the Laplacian matrix of the graph have been exploited, e.g., to estimate the connectivity of the network \cite{Fiedler}-\cite{Aragues-Shi-Dimarogonas-Sagues-Johansson}, to find densely connected clusters of nodes \cite{Bojan}-\cite{Hagen-Kaghn}, and to search for potential links that would greatly improve the connectivity if they were established \cite{Gosh-Boyd}. A recent tutorial work that gives an excellent view of this topic is \cite{Bertrand-Moonen2}. In all these works, it was shown that the connectivity properties of a graph can be assessed by looking at the second smallest eigenvalue of the Laplacian matrix, also known as the \textit{algebraic connectivity}, whereas a significant role in graph partitioning is plaid by the eigenvector associated to the algebraic connectivity, also known as \textit{Fiedler vector} \cite{Fiedler}. It is then important to find efficient algorithms to compute these connectivity parameters.
The problem of distributed estimation of the eigenvalues of the Laplacian matrix has been considered in several previous works, e.g., \cite{Kempe-McSherry}-\cite{Bertrand-Moonen2}. In \cite{Kempe-McSherry} a distributed algorithm is proposed to find the $n$ eigenvectors corresponding to the $n$ largest eigenvalues of the Laplacian matrix or the (weighted) adjacency matrix, based on power iteration and random walk techniques. The work in \cite{Franceschelli-Gasparri-Giua-C. Seatzu} evaluates the eigenstructure of the Laplacian matrix by letting the nodes oscillate at the eigenfrequencies corresponding to the network topology. A distributed algorithm that computes the Fiedler vector and the algebraic connectivity, with application to topology inference in ad hoc networks, has been also proposed in \cite{Bertrand-Moonen}-\cite{Bertrand-Moonen2}. The algorithms in \cite{Franceschelli-Gasparri-Giua-C. Seatzu}-\cite{Bertrand-Moonen2} allow to estimate the algebraic connectivity and the Fiedler vector without the need for nested consensus iterations, thus significantly reducing the communication burden. Distributed methods aimed at controlling the algebraic connectivity for flocking maintenance have also been proposed in   \cite{Yang-Freeman}-\cite{Zavlanos-Egerstedt-Pappas}. In \cite{Yang-Freeman}, the authors propose a distributed algorithm that allows each node to estimate and track the algebraic connectivity of the graph in mobile wireless sensor networks. Then, based on this estimator, a decentralized gradient controller for each agent helps maintain global connectivity during motion. The work in \cite{Zavlanos-Tanner-Jadbabaie-Pappas} enforces network connectivity by means of distributed topology control that decides on both deletion and creation of communication links between agents. With this protocol ensuring network connectivity, a decentralized motion controller aligns agent velocity vectors and regulates inter-agent distances to maintain existing network links. In \cite{Zavlanos-Egerstedt-Pappas}, a theoretical framework for controlling graph connectivity in mobile robot networks is given, thus providing centralized and distributed algorithms to maintain, increase, and control connectivity. Finally, references \cite{Montijano-Montijano-Sagues}-\cite{Bertrand-Moonen3} propose distributed methods for estimating the algebraic connectivity with the aim of optimizing the performance of average consensus algorithms. In \cite{Montijano-Montijano-Sagues} a consensus method using Chebyshev polynomials is considered and the authors proposed a distributed algorithm to compute the parameters that enable the method to maximize the convergence rate. Then, the work in \cite{Aragues-Shi-Dimarogonas-Sagues-Johansson} proposes a distributed algorithm to estimate the algebraic connectivity of a graph, thus applying this method to an event-triggered consensus scenario, where the most recent estimate of the algebraic connectivity is used for adapting the behavior of the average consensus algorithm. Finally, in \cite{Bertrand-Moonen3} the authors proposed a topology-aware distributed algorithm for on-line adaptation of the Laplacian weighting rule, when applied in an in-network averaging procedure.

All these previous works assumed ideal communications among the network nodes. However, in a realistic scenario, the wireless channels are affected by random fading and additive noise, which induce errors in the received packets. Furthermore, realistic random MAC protocols may determine packet collisions during the exchange of data among the nodes. In such a case, the receiving node could request the retransmission of the erroneous packets, but this would imply random delays in the communication among the nodes and it would be complicated to implement over a totally decentralized system. It is then of interest to analyze networks where the erroneous packets are simply dropped, without requiring a retransmission. Random packet dropping can be modeled as having random switching graph topologies. The effect of random graphs on distributed algorithms has been thoroughly studied in a series of works, mainly focused on the convergence of consensus algorithms, e.g., \cite{Hatano}-\cite{Kar-Moura2}, and of swarming algorithms for radio resource allocation \cite{DiLo_Barb2}.


In this work, we propose a distributed algorithm, based on a stochastic power iteration method, whose aim is to estimate the algebraic connectivity and the related Fiedler vector of the expected Laplacian matrix of a random graph, incorporating random impairments in the exchange of data among the nodes. The basic contributions of this paper are the following: 1) a novel algorithm to estimate in a distributed fashion the spectral connectivity parameters of the expected Laplacian matrix of a random graph; 2) the derivation of the convergence properties of the proposed algorithm in the presence of random link failures in the communications among nodes; 3) the control of the expected connectivity through the adaptation of the power transmitted by each node, in order to maximize the network connectivity in the presence of realistic MAC protocols or simply to drive it toward a desired target value.

The paper is organized as follows. In section II we first recall some basic concepts from algebraic graph theory that will be used throughout the paper. Then, we describe the proposed stochastic power iteration method for estimating the connectivity of a random graph, thus illustrating the distributed implementation based on consensus algorithms to decentralize the computation. The convergence properties of the proposed algorithm in the presence of random link failures are also investigated. In Section III, exploiting the proposed strategy for connectivity estimation, we propose a simple power control method aimed at controlling the expected connectivity of a network. Section IV then shows the effect of collisions, which are induced by a realistic random medium access control protocol, on the network connectivity. In particular, it is shown how, by choosing a too large transmission power, the network connectivity may be heavily degraded due to an increase in the collision probability. Then, we propose a distributed algorithm to evaluate the optimal transmission power that maximizes the connectivity in the presence of realistic MAC protocols. Finally, Section V draws some conclusions.

\section{Estimation of Algebraic Connectivity over Random Graphs}

\subsection{Algebraic Graph Theory}

We consider a network composed of $N$ nodes interacting according to a communication topology. The interaction among the nodes is modeled as an undirected graph $G=(V,E)$, where  $V={1,2,...,N}$ denotes the set of nodes and $E\subseteq V\times V$ is the edge set. The structure of the graph is described by a symmetric $N\times N$ adjacency matrix $\bA:=\{a_{ij}\}$, whose entries $a_{ij}$ are either positive or zero, depending on wether there is a link between nodes $i$ and $j$ or not, i.e., if the distance between nodes $i$ and $j$ is less than a coverage radius, which is dictated by nodes' transmit power and the channel between them. The set of neighbors of a node $i$ is ${\cal N}_i$, defined as ${\cal N}_i=\{j\in V:a_{ij}>0\}$. Node $i$ communicates with node $j$ if $j$ is a neighbor of $i$ (or $a_{ij}>0$). The graph has no self loops, i.e., $a_{ii}=0$ for all $i$. Denoting by $d_{ii}=\sum_{j=1}^{N}a_{ij}$ the degree of node $i$,  the degree matrix $\bD$ is a diagonal matrix with entries $d_{ii}$ that are the row sums of the adjacency matrix $\bA$. The graph Laplacian $\bL$ is an $N\times N$ matrix defined as
\begin{equation}
\bL=\bD-\bA.
\end{equation}
The spectral properties of $\bL$ have been shown to be critical in many multiagent applications, such as formation control \cite{Fax_murray}, consensus seeking \cite{Barb-Scut} and direction alignment \cite{Jad-Lin-Morse}. We denote by $\lambda_i(\bL)$, $i=1, \ldots, N$, the eigenvalues of $\bL$, ordered in increasing sense. The matrix $\bL$ always has, by construction, a null eigenvalue $\lambda_1(\bL)=0$, with associated eigenvector $\mathbf{1}$ composed of all ones. For a connected graph, the nullspace of $\bL$ has dimension $1$ and it is spanned by the vector $\mathbf{1}$. The second smallest eigenvalue $\lambda_2(\bL)$ is known as the {\it algebraic connectivity} of the graph. This eigenvalue is greater than 0 if and only if $G$ is a connected graph. The magnitude of this value reflects how well connected the overall graph is. For this reason, it has been used for example in analysing the synchronizability of networks \cite{Olfati1}-\cite{Scutari-Barbarossa-Pescosolido}, \cite{Fax_murray}-\cite{Barb-Scut}, in maintaining stable flocking   \cite{Jad-Lin-Morse}, and for routing optimization in cognitive radio ad-hoc networks \cite{Abb-Cuo}.

\noindent{\textit{Random link failures:}} In a realistic communication scenario, the packets exchanged among the nodes may be received with errors, because of collisions, channel fading or noise. The retransmission of erroneous packets can be incorporated into the system, but packet retransmission introduces a nontrivial additional complexity in decentralized implementations and, more importantly, it also introduces an unknown delay and delay jitter. It is then of interest to examine protocols where erroneous packets are simply dropped. We take into account random packet dropping by modeling the coefficient $a_{ij}$ describing the network topology as statistically independent random variables. Then, the Laplacian of the graph varies with time as a sequence of i.i.d. matrices $\{\bL[k]\}$, which can be written, without any loss of generality, as
\begin{equation}\label{RandomLaplacian}
 \bL[k] =\bar{\bL} + \tilde{\bL}[k]
\end{equation}
where $\bar{\bL}=\{\bar{l}_{ij}\}$ denotes the expected matrix and $\tilde{\bL}[k]=\{\tilde{l}_{ij}[k]\}$ are i.i.d. perturbations around the mean.
The i.i.d. fluctuations $\tilde{l}_{ij}[k]$ affect only the active links, i.e. the links for which $a_{ij}\neq 0$; for all other inactive links, the perturbations are equal to zero. We do not make any assumptions of symmetry of the failures, i.e. $\tilde{l}_{ij}[k]$ may be not equal to $\tilde{l}_{ji}[k]$, or about the link failure model. Although the link failures are independent over time, during the same iteration, the link failures can still be spatially correlated. It is important to remark that, in the ensuing analysis and derivations, we do not require the random instantiations $G[k]$ of the graph be connected for all $k$. We only require the graph to be connected on average. This condition is captured by requiring $\lambda_2(\bar{\bL})>0$.

\subsection{Stochastic Power Iteration}

In this section, we propose a novel algorithm aimed at assessing the connectivity of a random graph by estimating the second smallest eigenvalue of the expected Laplacian matrix $\bar{\bL}$. Since in our setting the network graph is random due to the presence of link failures, we introduce a stochastic power iteration method to handle the randomness introduced by the graph fluctuation.

Let us introduce the matrix $\bW[k]$ given at time $k$ by:
\begin{equation}\label{Consensus_matrix}
\bW[k]=\bI-\bar{\varepsilon}\bL[k]=\bar{\bW}+\tilde{\bW}[k]
\end{equation}
where $\bar{\varepsilon}$ is a positive parameter, $\bar{\bW}=\bI-\bar{\varepsilon}\bar{\bL}$ is the mean matrix, and $\tilde{\bW}[k]=-\bar{\varepsilon}\tilde{\bL}[k]$ are i.i.d. fluctuations around the mean.
The matrix $\bW[k]$ in (\ref{Consensus_matrix}) was used, for example, as the iteration matrix of consensus algorithms over random graphs, see e.g. \cite{Tahbaz-Salehi-Jadb1}-\cite{Tahbaz-Salehi-Jadb2}, \cite{Kar-Moura}-\cite{Kar-Moura2}.
From (\ref{Consensus_matrix}), the eigenvalues of the expected Laplacian matrix $\bar{\bL}$ are directly related to those of the expected matrix $\bar{\bW}$ in (\ref{Consensus_matrix}) through the relation
\begin{equation}\label{eigenv_Lapl_cons}
\lambda_i(\bar{\bL})=\frac{1-\lambda_{N+1-i}(\bar{\bW})}{\bar{\varepsilon}}, \quad i=1,\ldots,N.
\end{equation}
In particular, the algebraic connectivity is given by $\lambda_2(\bar{\bL})=(1-\lambda_{N-1}(\bar{\bW}))/\bar{\varepsilon}$. Furthermore, the eigenvector $\bu_{N-1}(\bar{\bW})$ associated to the second largest eigenvalue of the expected matrix $\bar{\bW}$ coincides with  $\bu_{2}(\bar{\bL})$, which is the one associated to the second smallest eigenvalue of the expected Laplacian matrix $\bar{\bL}$, also known as the Fiedler vector.
The coefficient $\bar{\varepsilon}$ in (\ref{Consensus_matrix}) satisfies
\begin{eqnarray}\label{Condition}
\quad 0<\bar{\varepsilon}<\frac{2}{\lambda_N(\bar{\bL})},
\end{eqnarray}
which, combined with the condition $\lambda_2(\bar{\bL})>0$, ensures that the mean matrix $\bar{\bW}$ is a Perron matrix having a single unitary eigenvalue \cite{Olfati1}. Since we want to track the second largest eigenvalue of the mean matrix $\bar{\bW}$, we deflate the original matrix $\bW[k]$ by removing its largest eigenvalue, thus obtaining the matrix $\bB[k]$ given by:
\begin{align}\label{Defl_matrix2}
\bB[k] =\bW[k]-\frac{1}{N}\mathbf{1}\mathbf{1}^T:=\bar{\bB}+\tilde{\bB}[k]
\end{align}
where $\displaystyle \bar{\bB}=\bar{\bW}-\frac{1}{N}\mathbf{1}\mathbf{1}^T$ and $\tilde{\bB}[k]=\tilde{\bW}[k]=-\bar{\varepsilon}\tilde{\bL}[k]$. In this way, the maximum eigenvalue of the deflated expected matrix $\bar{\bB}$ coincides with the second largest eigenvalue of $\bar{\bW}$.
To handle the randomness of the graph, we introduce also the deflated matrix
\begin{align}\label{Defl_matrix3}
\bB_2[k]=\bI-\varepsilon[k]\bL[k]-\frac{1}{N}\mathbf{1}\mathbf{1}^T=\bW_2[k]-\frac{1}{N}\mathbf{1}\mathbf{1}^T,
\end{align}
where $\bW_2[k]=\bI-\varepsilon[k]\bL[k]$, with $\varepsilon[k]$ denoting a positive diminishing sequence that we will choose in the sequel. The matrices $\bB_2[k]$ and $\bB[k]$ have exactly the same eigenvectors, but different eigenvalues,
at each time $k$, due to the time-varying sequence $\varepsilon[k]$.

We consider first a centralized implementation of the stochastic power iteration algorithm, whose main steps
are listed in Table 1. A distributed implementation of the algorithm will be illustrated later on.
\begin{algorithm}
\caption*{\textbf{Table 1: Centralized Stochastic Power Iteration}}

\vspace{.2cm}
Initialize $\bx[0]$, $y[0]$, and $z[0]$ randomly. Then, set $k=0$ and perform the following steps:
\begin{enumerate}
  \item Build the deflated matrices
        \begin{align}
               \bB[k]&=\bI-\bar{\varepsilon}\bL[k]-\frac{1}{N}\mathbf{1}\mathbf{1}^T\\
               \bB_2[k]&=\bI-\varepsilon[k]\bL[k]-\frac{1}{N}\mathbf{1}\mathbf{1}^T
        \end{align}
        where $\bar{\varepsilon}$ and $\varepsilon[k]$ satisfy (\ref{Condition}) and (\ref{epsilonk}), respectively;
  \item Evaluate the estimate $y[k+1]$ of $\lambda_{N-1}(\bar{\bW})$ as:
        \begin{align}
               y_0[k]&=\frac{\bx^T[k]\bB[k]\bx[k]}{\bx^T[k]\bx[k]}\label{lambdaN-1_2}\\
               y[k+1]&=y[k]+\alpha[k]\left(y_0[k]-y[k]\right)\label{lambdaN-1}
        \end{align}
        where $\alpha[k]$ is a time varying step-size satisfying (\ref{Step_size});
  \item Compute the estimate $z[k+1]$ of $\lambda_2(\bar{\bL})$ as:
       \begin{equation}\label{lambda2_est}
           z[k+1]=\frac{1-y[k+1]}{\bar{\varepsilon}};
       \end{equation}
  \item Perform the power iteration step
        \begin{equation}\label{pow_iter}
           \displaystyle \bx[k+1]=\frac{\bB_2[k]\bx[k]}{\|\bB_2[k]\bx[k]\|};
        \end{equation}
  \item If convergence is achieved stop, otherwise set $k=k+1$ and go to step 1.
\end{enumerate}
\end{algorithm}

The aim of the stochastic power iteration steps in (\ref{lambdaN-1_2})-(\ref{pow_iter}) is to estimate the largest eigenvalue of the expected matrix $\bar{\bB}$ (i.e. $\lambda_{N-1}(\bar{\bW})$), which is directly related to the second eigenvalue $\lambda_2(\bar{\bL})$ of the expected Laplacian through (\ref{lambda2_est}). To achieve this goal, the power iteration in (\ref{pow_iter}) runs over the matrix $\bB_2[k]$, thus providing an estimate $\bx[k]$ of the eigenvector $\bu_2(\bar{\bL})$ associated to the largest eigenvalue of the expected matrix $\bar{\bB}$ (remember that $\bB_2[k]$ and $\bB[k]$ have the same eigenvectors).
The eigenvector estimate $\bx[k]$ is then used in the Rayleigh ratio (\ref{lambdaN-1_2}) to provide an estimate $y_0[k]$ for $\lambda_{N-1}(\bar{\bW})$.
In both adaptations, it is fundamental to choose the step sizes $\varepsilon[k]$ in (\ref{Defl_matrix3}) and $\alpha[k]$ in (\ref{lambdaN-1}).
In particular,  we make the following assumptions, which are standard in stochastic approximation and adaptive signal processing \cite{Sayed}-\cite{Nevel}:

\noindent {\textit{Assumption A.1} :} (\emph{Persistence}) The sequence $\varepsilon[k]$ in (\ref{Defl_matrix3}) and the step-size sequence $\alpha[k]$ in (\ref{lambdaN-1}) satisfy the conditions
\begin{align}
&\varepsilon[k]>0, \hspace{.4cm}  \sum_{k=0}^{\infty}\varepsilon[k]=\infty, \hspace{.4cm}   \sum_{k=0}^{\infty}\varepsilon^2[k]<\infty,\label{epsilonk}\\
&\alpha[k]>0, \hspace{.4cm}  \sum_{k=0}^{\infty}\alpha[k]=\infty, \hspace{.4cm}   \sum_{k=0}^{\infty}\alpha^2[k]<\infty.\label{Step_size}
\end{align}
Conditions (\ref{epsilonk})-(\ref{Step_size})  ensure that the step size sequences decay to zero, but not too fast. An example of sequences satisfying (\ref{epsilonk})-(\ref{Step_size}) is
\begin{eqnarray}\label{Step_size2}
\varepsilon[k]=\frac{\varepsilon_0}{(k+1)^{\gamma}}, \quad\quad \alpha[k]=\frac{\alpha_0}{(k+1)^{\beta}},
\end{eqnarray}
$\varepsilon_0,\alpha_0>0, 0.5<\beta,\gamma\leq1.$

\noindent We are now able to state the main theorem on the convergence of the proposed stochastic power iteration method.

\noindent {\textit{Theorem 1} :} Let $z[k]$ and $\bx[k]$ be the sequences generated in (\ref{lambda2_est}) and (\ref{pow_iter}) by the stochastic power iteration. If $\lambda_{2}(\bar{\bL})>0$, and condition (\ref{Condition}) and Assumption A.1 hold, we have
\begin{align}
\lim_{k\rightarrow\infty} z[k]=\lambda_{2}(\bar{\bL}), \quad\hbox{and}\quad \lim_{k\rightarrow\infty} \bx[k]=\hat{\bu}_{2}(\bar{\bL}),
\end{align}
almost surely (w.p.1), where $\hat{\bu}_{2}(\bar{\bL})$ denotes the normalized Fiedler vector of the expected Laplacian matrix $\bar{\bL}$.

\begin{proof}
See Appendix B.
\end{proof}

Theorem 1 establishes the almost sure convergence of the stochastic power iteration method to the Fiedler vector $\hat{\bu}_{2}(\bar{\bL})$ and to the algebraic connectivity $\lambda_2(\bar{\bL})$. As shown in Appendix C, an upper bound on the slowest (undesired) decaying mode $c_r[k]$ of the algorithm is given by
\begin{align}\label{conv_mean_part2}
c_r[k]&\leq \exp \left(-\Big(\lambda_3(\bar{\bL})-\lambda_2(\bar{\bL})\Big)\sum_{l=0}^k\varepsilon[l]\right).
\end{align}
From (\ref{conv_mean_part2}), we see how the convergence rate depends on the difference between the third and the second eigenvalues of the expected Laplacian $\bar{\bL}$, and on the sequence $\varepsilon[k]$. In particular, (\ref{conv_mean_part2}) makes clear that, because of (\ref{epsilonk}), the slowest decaying mode goes to zero as $k\rightarrow\infty$.

\noindent{\it Remark:} As mentioned in the introduction, the eigenvectors of the Laplacian matrix give useful information about how the network can be partitioned, i.e., how to find clusters of nodes in the network. It has been shown in several works, e.g., \cite{Bojan}-\cite{Hagen-Kaghn}, that spectral clustering can infer more topological properties of the graph if more eigenvectors of the Laplacian matrix, besides the one associated to the algebraic connectivity, are known. Up to now, it was shown how the proposed stochastic power iteration can compute only the second smallest eigenvalue and the corresponding eigenvector of the expected Laplacian matrix of the graph. However, the trick of deflating the matrix $\bW[k]$ in (\ref{Defl_matrix2}) can be sequentially iterated in order to estimate the third order eigenvalue and the associated eigenvector.
As an example, once the algebraic connectivity $\lambda_{2}(\bar{\bL})$ and the Fiedler vector $\bu_{2}(\bar{\bL})$ have been estimated through a first stage power iteration, the third order eigenparameters can be estimated by applying again power iteration using the pair of deflated matrices
\begin{align}
\bC[k]&=\bB[k]-(1-\bar{\varepsilon} \lambda_{2}(\bar{\bL}))\bu_{2}(\bar{\bL})\bu^T_{2}(\bar{\bL}),\nonumber\\
\bC_2[k]&=\bB_2[k]-(1-\varepsilon[k] \lambda_{2}(\bar{\bL}))\bu_{2}(\bar{\bL})\bu^T_{2}(\bar{\bL}), \nonumber
\end{align}
which take the role of $\bB[k]$ and $\bB_2[k]$ in Table 1, respectively. In a similar way, we can estimate also the higher order eigen-parameters by sequential deflation and power iteration.

\noindent{\emph{Distributed implementation :}} The stochastic power iteration method described before requires a centralized implementation. In the following, we propose a decentralized implementation based on average consensus \cite{Olfati1},\cite{Barb-Scut}. The two operations to be distributed are the Rayleigh ratio in (\ref{lambdaN-1_2}) and the power iteration in (\ref{pow_iter}), whereas all other computations can be performed locally. 
Setting $\bb[k]=\bB[k]\bx[k]$ and $\bb_2[k]=\bB_2[k]\bx[k]$, the $i$-th components $b_i[k]$ and $b_{2,i}[k]$ of the vectors $\bb[k]$ and $\bb_2[k]$ can be evaluated locally. In fact, exploiting the structure of the matrices in (\ref{Defl_matrix2}) and (\ref{Defl_matrix3}), we have
\begin{eqnarray}
    b_i[k]=x_i[k]+\bar{\varepsilon}\sum_{j=1}^Na_{ij}[k](x_j[k]-x_i[k])-m[k]\label{b_i}\\
    b_{2,i}[k]=x_i[k]+\varepsilon[k]\sum_{j=1}^Na_{ij}[k](x_j[k]-x_i[k])-m[k]\label{b_i2}
\end{eqnarray}
where $\displaystyle m[k]=\frac{1}{N}\mathbf{1}^T\bx[k]$ is a global parameter.
The value $m[k]$ is given by the average of the values $x_i[k]$ stored locally at each node, and can be computed in a decentralized fashion using a round of average consensus protocol. The next step is to evaluate the Rayleigh ratio in (\ref{lambdaN-1_2}) in a distributed fashion. To this end, we notice that expression (\ref{lambdaN-1_2}) can be recast as
\begin{eqnarray}\label{lambdaN-1_3}
\displaystyle\frac{\bx^T[k]\bB[k]\bx[k]}{\bx^T[k]\bx[k]}
=\frac{\sum_{i=1}^N x_i[k]b_i[k]}{\sum_{i=1}^N x^2_i[k]},
\end{eqnarray}
where both numerator and denominator are written as inner products. This notation is convenient because it enables us to compute this expression through a step of weighted average consensus \cite{Olfati1},\cite{Barb-Scut}, which evaluates in a distributed manner the ratio in (\ref{lambdaN-1_3}). Thus, at this stage, each node is able to compute (\ref{lambdaN-1_2}) and (\ref{lambdaN-1}) locally.
To complete the series of operations of the stochastic power iteration algorithm, node $i$ still needs  to evaluate (\ref{pow_iter}) in a distributed fashion. Then, each node $i$ computes the $i$th component of vector $\bx[k+1]$ in (\ref{pow_iter}) as:
\begin{equation}\label{pow_iter3}
     x_i[k+1]=\frac{b_{2,i}[k]}{\|\bb_2[k]\|}.
\end{equation}
Since the numerator has been already computed through (\ref{b_i2}), we only need to compute the denominator of (\ref{pow_iter3}). In particular, we consider the evaluation of
\begin{equation}\label{scaled_norm}
\frac{1}{\sqrt{N}}\|\bb_2[k]\|=\sqrt{\frac{1}{N}\sum_{i=1}^N b^2_{2,i}[k]},
\end{equation}
which is a scaled version of $\|\bb_2[k]\|$, and can be computed in a distributed fashion by taking the square root of the output of an average consensus step. Each node then computes $\displaystyle \hat{x}_i[k+1]=\sqrt{N}\frac{b_{2,i}[k]}{\|\bb_2[k]\|}$,
which is a scaled version of the true value $x_i[k]$ that the algorithm should compute in (\ref{pow_iter3}). However, even in the presence of such update, the method still works correctly because, at time $k+1$, the step in (\ref{lambdaN-1_2}) is a Rayleigh ratio, whose result is not affected by the scaling $\sqrt{N}$, thus leading to the correct update of the algorithm.

The main steps of the decentralized implementation are summarized in Table 2. We also define $\delta$-convergence of a sequence $c[k]$ the event $\displaystyle\frac{|c[k+1]-c[k]|}{|c[k]|}\leq\delta$.

\begin{algorithm}
\caption*{\textbf{Table 2: Distributed Stochastic Power Iteration}}

\vspace{.2cm}
Each node initializes $x_i[0]$, $y[0]$, and $z[0]$ randomly. Then, set $k=0$ and performs the following steps:

\begin{enumerate}
  \item Run a consensus round to get $\displaystyle m[k]=\mathbf{1}^T\bx[k]/N$ until $\delta_1$-convergence;
  \item Evaluate $b_i[k]$ and $b_{2,i}[k]$, $\forall i$, using (\ref{b_i})-(\ref{b_i2});
  \item Run a consensus round to compute the Rayleigh ratio in (\ref{lambdaN-1_3}) (i.e. $y_0[k]$ in (\ref{lambdaN-1_2})) and the scaled norm in (\ref{scaled_norm}) until $\delta_2$-convergence;
  \item Compute the estimate $y[k+1]$ of $\lambda_{N-1}(\bar{\bW})$ using (\ref{lambdaN-1});
  \item Compute the estimate $z[k+1]$ of $\lambda_{2}(\bar{\bL})$ using (\ref{lambda2_est});
  \item Perform the power iteration in (\ref{pow_iter3}), $\forall i$.
  \item If convergence is achieved stop, otherwise set $k=k+1$ and go to step 1.
\end{enumerate}
\end{algorithm}

\noindent{\it Remark 2 :} The distributed implementation of the stochastic power iteration proposed in Table 2 is based on the use of two rounds of average consensus algorithm \cite{Olfati1},\cite{Barb-Scut}, which allow the computation of the global quantities in the steps 1 and 3 in Table 2. Of course, also the consensus algorithm will be affected by the presence of random link failures.
However, if the expected graph is connected (i.e., $\lambda_2(\bar{\bL})>0$) and the matrices $\bW[k]$ in (\ref{Consensus_matrix}) are doubly stochastic for every $k$, it is well known that consensus algorithm is robust to the presence of link failures \cite{Tahbaz-Salehi-Jadb1}-\cite{Kar-Moura2}, thus guaranteing convergence to the desired average value. Thus, differently from the centralized case in Table 1, where no symmetry assumptions are required on the link failures, in the distributed implementation the network graph must be balanced at every iteration.
Regarding the communication demands of the proposed distributed implementation, in the first consensus round each node must broadcast a scalar value to its neighbors, whereas in the second round it is necessary to transmit two scalar values. Thus, letting $C$ be the cost associated to the transmission of a scalar value, the communication cost per iteration $k$ of the power iteration algorithm is equal to $(N_1+2N_2)C$, where $N_1$ and $N_2$ are the number of iterations needed by the first and second consensus rounds to converge within  prescribed accuracies $\delta_1$ and $\delta_2$, respectively. The communication demand is then determined by the convergence rate of the consensus algorithm, which depends on the value of $\lambda_2(\bar{\bL})$, as shown in several previous works, see, e.g., \cite{Kar-Moura}, \cite{Kar-Moura2}. In particular, the more connected is the expected graph (i.e., larger values of $\lambda_2(\bar{\bL})$) the faster is convergence of consensus, as illustrated in the following numerical examples.

\begin{figure}[t]
\centering
\includegraphics[width=7cm]{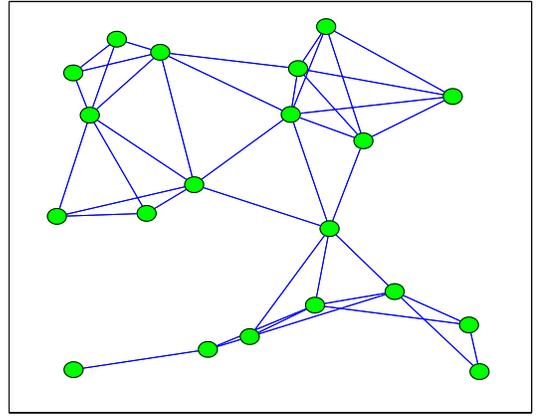}
  \caption{Network topology}\label{Net}
\end{figure}

\begin{figure}[t]
\centering
\includegraphics[width=8.5cm]{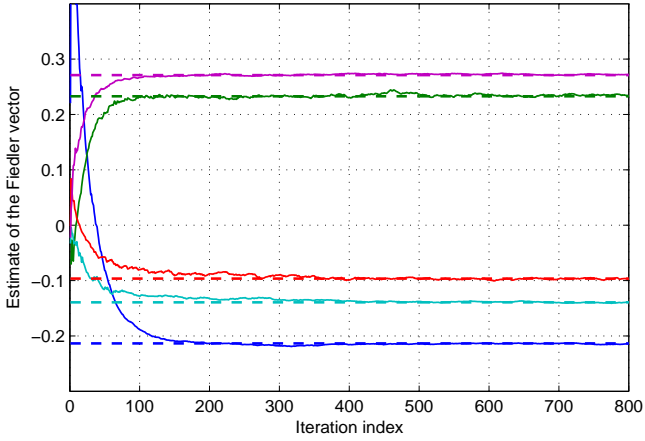}
  \caption{Estimate of the Fiedler vector versus iteration index}\label{Fiedler}
\end{figure}

\begin{figure}[t]
\centering
\includegraphics[width=8.5cm]{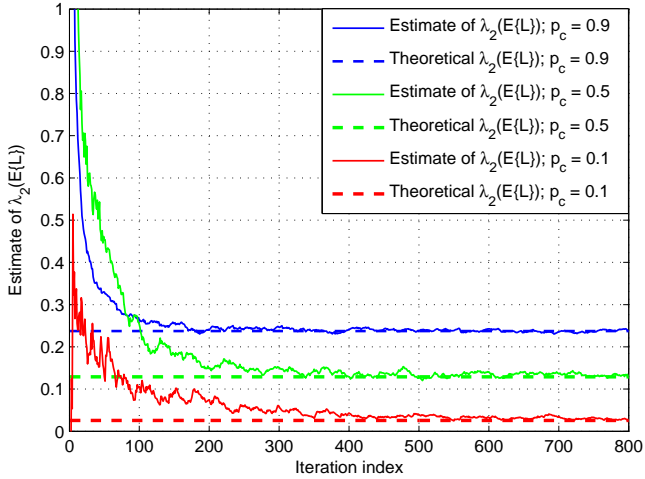}
  \caption{Estimate of $\lambda_2(\bar{\bL})$ versus iteration index, for different probabilities to establish a communication link.}\label{lambda2}
\end{figure}

\begin{figure}[t]
\centering
\includegraphics[width=8.1cm]{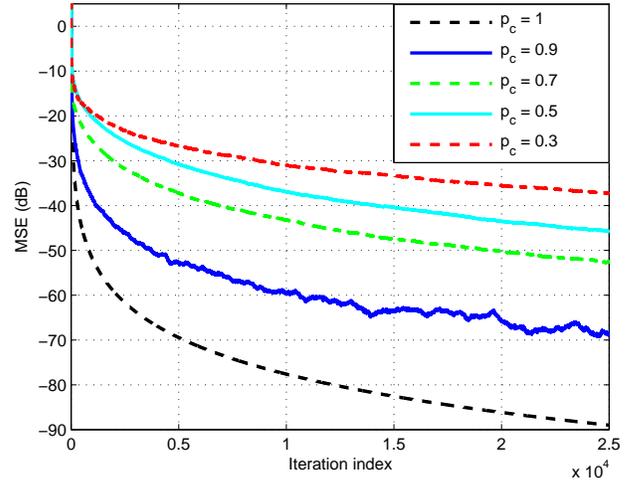}
  \caption{MSE versus iteration index, for different probabilities to establish a communication link.}\label{MSE_lambda2}
\end{figure}

\noindent{\it Numerical example - Convergence:} The aim of this example is to corroborate the theoretical results in Theorem 1, which establishes the almost sure convergence of the stochastic power iteration method in Table 1. We consider a connected network composed of 20 nodes, whose ideal topology (in the absence of link failures) is shown in Fig. \ref{Net}. The communication among nodes is impaired by random link failures so that each link in Fig. \ref{Net} is on with a certain probability $p_c$, which is here assumed to be constant over all links.
In Fig. \ref{Fiedler}, we report the behavior of the estimate of five components of the Fiedler vector  $\bu_2(\bar{\bL})$ versus the iteration index, obtained setting $p_c=0.8$. The theoretical value of each component is also reported as a horizontal dashed line. The sequences $\alpha[k]$ and $\varepsilon[k]$ are chosen as in (\ref{Step_size2}), with $\alpha_0=1.5$, $\beta=0.51$, $\varepsilon_0=0.4$, $\gamma=0.51$, in order to satisfy (\ref{Step_size}). As we can notice from Fig. \ref{Fiedler}, the algorithm asymptotically converges to the theoretical value of Fiedler vector, thus confirming the theoretical results obtained in Theorem 1.
Moreover, in Fig. \ref{lambda2}, we report the behavior of the estimate of $\lambda_2(\bar{\bL})$ versus the iteration index, considering different probabilities to establish a communication link. The theoretical values of $\lambda_2(\bar{\bL})$ are also reported as horizontal dashed lines. The parameters are the same as in the previous simulation. As we can notice from Fig. \ref{lambda2}, the algorithm converges to the theoretical value of the algebraic connectivity.
To validate the almost sure convergence claimed in Theorem 1 numerically, in Fig. \ref{MSE_lambda2}, we report the behavior of the mean square error (MSE) on the estimate of $\lambda_2(\bar{\bL})$, i.e. $\mathbb{E}\{(z[k]-\lambda_2(\bar{\bL}))^2\}$, considering different probabilities to establish a communication link. The results are averaged over 100 independent realizations. The parameters are chosen as $\alpha_0=1.5$, $\beta=0.9$, $\varepsilon_0=0.6$, $\gamma=0.9$. The ideal case, corresponding to $p_c=1$, is also reported as a benchmark. From Fig. \ref{MSE_lambda2}, we notice how the MSE goes to zero as $k\rightarrow\infty$, for any value of the probability $p_c$ to establish a link. As expected, we can also see how, reducing the probability to establish a communication link, the convergence rate of the algorithm decreases.

\begin{figure}[t]
\centering
\includegraphics[width=8.1cm]{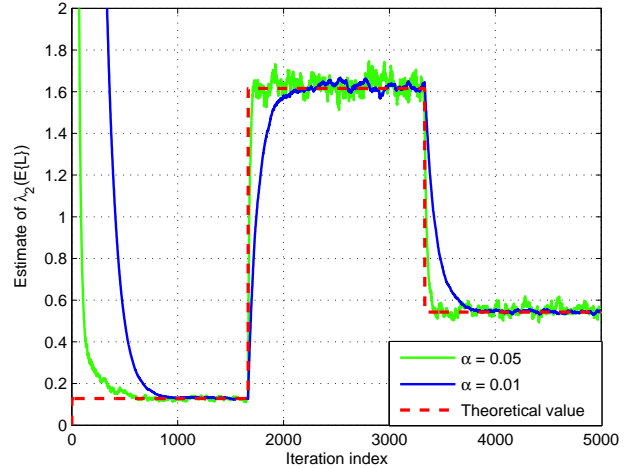}
  \caption{Adaptive estimate of $\lambda_2(\bar{\bL})$ versus iteration index.}\label{tracking}
\end{figure}

\noindent{\it Numerical example - Adaptation to time-varying scenarios:} In many practical applications, the connectivity may change over time due to many factors like, e.g., nodes' mobility, nodes' failures, variation of channel conditions, etc. It is then of interest to devise adaptive techniques that are able to track the temporal variation of the expected graph's connectivity. The stochastic power iteration in (\ref{lambdaN-1_2})-(\ref{pow_iter}) converges to the desired eigenvalue almost surely thanks to the effect of the diminishing sequences in (\ref{Step_size})-(\ref{Step_size2}), which asymptotically drives to zero the noise variance. However, this method is not adaptive since the use of a diminishing step-size would make impossible a short-term adaptation to temporal variations of the expected graph connectivity. It is then of interest to check the tracking capabilities of the  algorithm, assuming constant step sizes, i.e. $\alpha[k]=\alpha_0$ in (\ref{lambdaN-1}) and $\varepsilon[k]=\bar{\varepsilon}$ in (\ref{Defl_matrix3}). To assess the adaptation capability of the proposed method to temporal changes in the algebraic connectivity of the network expected graph, we consider a scenario where the ideal connectivity of the graph varies with time between three different values. The probability to establish a communication link is kept fixed at $p_c=0.5$, and $\bar{\varepsilon}=0.1$. In Fig. \ref{tracking} we illustrate the behavior of the estimate of $\lambda_2(\bar{\bL})$ versus the iteration index, considering the adaptive stochastic power iteration method with two different constant step-size values $\alpha$. The theoretical value of $\lambda_2(\bar{\bL})$ is also reported to illustrate the convergence properties. As we can see from Fig. \ref{tracking}, the adaptive implementation allows online tracking of the algebraic connectivity of the expected graph. In particular, we can notice from Fig. \ref{tracking} how a larger step-size leads to a better adaptation capability, at the cost of larger variance of the estimation error.

\begin{figure}[t]
\centering
\includegraphics[width=8.3cm]{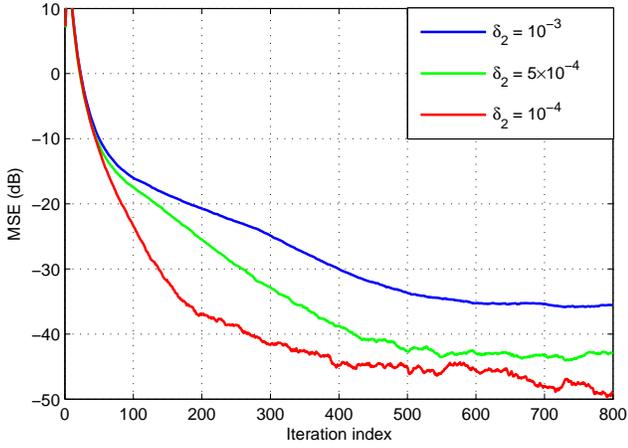}
  \caption{MSE versus iteration index, for different values of $\delta_2$.}\label{MSE_delta}
\end{figure}

\begin{figure}[t]
\centering
\includegraphics[width=8cm]{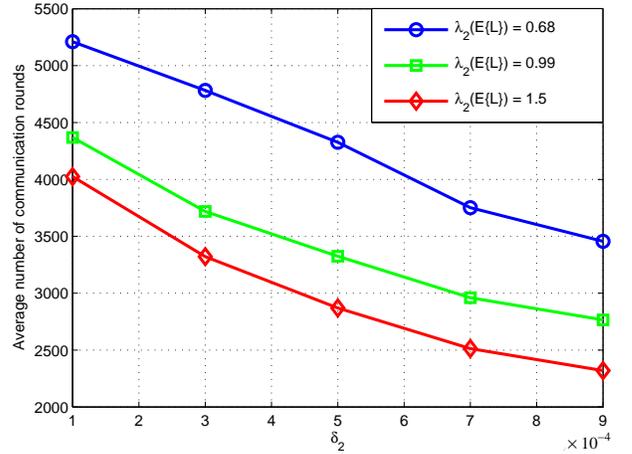}
  \caption{Average number of communication rounds versus $\delta_2$, for different values of $\lambda_2(\bar{\bL})$.}\label{Iter_delta}
\end{figure}

\noindent{\it Numerical example - Performance of the distributed implementation:} The distributed implementation of the stochastic power iteration method in Table 2 relies on two nested consensus loops per iteration. Each of them runs until it reaches a certain precision determined by the parameters $\delta_1$ and $\delta_2$. Since this premature stop inevitably introduces an approximation error in the evolution of the stochastic power iteration, it is of interest to evaluate the effect of such an error on the performance of the algorithm. Then, in Fig. \ref{MSE_delta}, we illustrate the behavior of the MSE on the estimate of $\lambda_2(\bar{\bL})$ versus the iteration index, achieved by the distributed stochastic power iteration algorithm in Table 2, considering different values of the approximation parameter $\delta_2$. The results are averaged over 100 independent simulations. The network topology is achieved from the one in Fig. \ref{Net} by increasing the algebraic connectivity of the ideal graph (with no failures) to $\lambda_2(\bL)=1.08$. The link failure probability is set to $p_c=0.9$, and the approximation parameter $\delta_1=0.1$. The sequences $\alpha[k]$ and $\varepsilon[k]$ are chosen as in (\ref{Step_size2}), where $\alpha_0=1$, $\beta=0.8$, $\varepsilon_0=0.35$, $\gamma=0.51$. From Fig. \ref{MSE_delta}, we notice how the MSE converges to a finite value due to the presence of a bias introduced by the approximation errors in the estimate of $\lambda_2(\bar{\bL})$. As expected, we can see how, using a smaller value of $\delta_2$ (higher precision in the final consensus value), the bias is reduced and the algorithm shows better performance in terms of MSE. However, this benefit does not come without a price as, reducing the value of the approximation parameter $\delta_2$, the inner consensus loops will require more iterations to converge with the desired accuracy. To give an example of the overall communication burden, in Fig. \ref{Iter_delta} we report the number of communication rounds, needed by the proposed distributed algorithm to converge, versus the approximation parameter $\delta_2$. We consider different values of the algebraic connectivity of the expected graph, which are obtained by varying the connectivity of the underlying ideal graph while keeping fixed the link failure probability to $p_c=0.9$. The behaviors are averaged over 200 independent simulations. To obtain these results, we have defined a convergence criterion also for the sequence $z[k]$ in (\ref{lambda2_est}). In particular, we consider a $\delta_3$-convergence criterion, with $\delta_3=5\times 10^{-4}$. As we can notice from Fig. \ref{Iter_delta}, reducing the value of $\delta_2$, the algorithm requires a larger communication burden to converge within the desired accuracy. Furthermore, increasing the algebraic connectivity of the expected graph, the overall number of communication decreases due to the improved convergence rate of the power iteration and the inner consensus loops. In summary, the choice of the precision parameters $\delta_1$, $\delta_2$, and $\delta_3$ introduces a tradeoff between achievable performance at convergence and communication burden of the algorithm.

\section{Connectivity Control}

The proposed algorithm can be applied to estimate the connectivity of all kinds of random networks. Assuming the nodes are deployed according to a random geometric graph (RGG) \cite{Penrose}, the stochastic power iteration method can be used to adapt the power transmitted by each node in order to drive the network connectivity toward a desired value. This can be obtained through a power control step, where each node updates its transmission power as:
\begin{equation}\label{power_update}
P_T[k+1]=P_T[k]+\mu\left(\lambda^*-z[k+1]\right)
\end{equation}
for $k\geq0$, where $\lambda^*$ is a positive constant used to enforce a desired connectivity value, $\mu$ is a positive step-size, and $z[k+1]$ is the estimate of $\lambda_2(\bar{\bL})$, at time $k+1$, carried out by the stochastic power iteration method in (\ref{lambda2_est}). Intuitively, the proposed controller increases the power transmitted by each node if the current estimated value of $\lambda_2(\bar{\bL})$ is lower than the desired value, whereas it reduces the transmitted power in the opposite case.
The power update in (\ref{power_update}) can be inserted as fifth step of the stochastic power iteration in (\ref{lambdaN-1_2})-(\ref{pow_iter}), thus leading to a dynamic change of the network topology toward a desired connectivity value.

\begin{figure}[t]
\centering
\includegraphics[width=8cm]{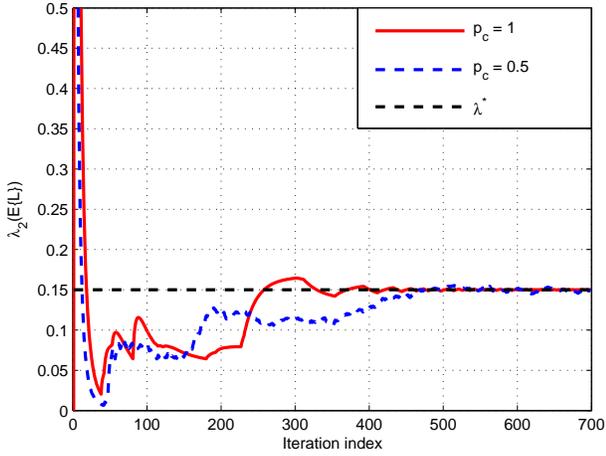}
  \caption{Behavior of $\lambda_2(\bar{\bL})$ versus iteration index.}\label{lambda2pc}
\end{figure}

\noindent{\it Numerical example :} In this example we combine the stochastic power iteration step in (\ref{lambdaN-1_2})-(\ref{pow_iter}) with the power control step in (\ref{power_update}), thus illustrating the capability of the resulting strategy to control the connectivity of the network random graph.
As a starting point, we consider a network composed of 25 nodes deployed over a geographic area of 1600 $m^2$ according to a certain initial topology having an initial value of algebraic connectivity $\lambda_2(\bL)=0.0599$. The initial power transmitted by each node is $p_i[0]=1$ mW, whereas the minimum received power needed to establish a communication link among two nodes is $P_{th}=0.01$ mW. We assume a free-space path loss as a propagation environment, i.e. $\xi=2$. Our goal is to use the proposed algorithm to drive the connectivity of the expected graph toward a desired value $\lambda^*=0.15$, considering two different values of probability to establish a link, e.g., $p_c=1$, and $p_c=0.5$. In Fig. \ref{lambda2pc}, we report the behavior of the estimate of $\lambda_2(\bar{\bL})$ in (\ref{lambda2_est}) versus the iteration index. The continuous curve illustrates the case $p_c=1$, whereas the dashed curve the case $p_c=0.5$. The step-sizes are chosen as in (\ref{Step_size2}), where $\alpha_0=1$, $\beta=0.55$, $\varepsilon_0=0.1$, $\gamma=0.55$, and $\mu=0.05$. As we can notice from Fig. \ref{lambda2pc}, the value of the algebraic connectivity of the expected graph converges very close to the desired value $\lambda^*$ for both values of $p_c$.
\begin{figure}[t]
\centering
\includegraphics[width=8cm]{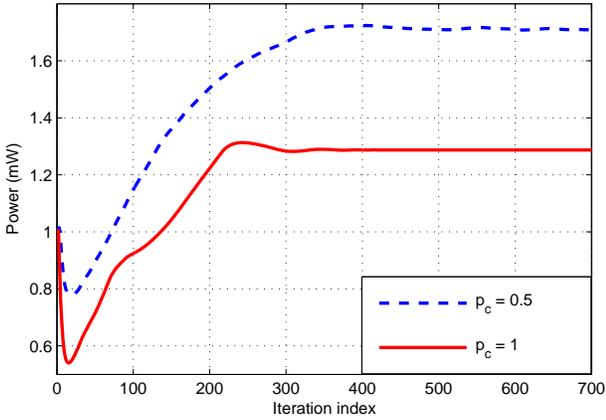}
  \caption{Temporal behavior of the power transmitted by nodes.}\label{sum_power}
\end{figure}
In Fig. \ref{sum_power}, we illustrate the temporal behavior of the power transmitted by each node. Form Fig. \ref{sum_power}, we can see how the power converges to a fixed value, which depends on the probability to establish a communication link. In particular, reducing the probability to establish a communication link with respect to the ideal case, the algorithm will increase the number of links of the resulting network in order to counteract the effect of failures and reach the target value of connectivity. Then, each node will transmit more power to enlarge its own subset of neighbors.

\section{Connectivity of wireless ad-hoc networks with realistic MAC}

In a realistic communication scenario, nodes communicate with each other by accessing to a shared channel according to a specified MAC protocol. Let us assume that, in the considered wireless ad-hoc scenario, each node has $M$ wireless channels that are dedicated to the exchange of data with its own neighbors. To establish a communication, a node randomly selects one of these channels independently of the choices of its neighbors. Let us further assume that the nodes are deployed according to an RGG. It is well known that asymptotically, as the number of nodes goes to infinity, RGG networks tend to satisfy a regularity condition, i.e., each node has the same number $d$ of neighbors on average \cite{Penrose}. The average number $d$ of neighbors depends on the covering radius of each node, which is dictated by the transmitted power and the channel conditions. Let us assume a simple free-space propagation model so that the power received by a node is related to the transmitted power as $P_R=P_{T}/r^2$, where $r$ is the covered distance. Now, setting a minimum threshold value $P_{th}$ for the power at the receiver node, the covering radius is just obtained by inverting the previous expression as $r^2=P_{T}/P_{th}$. The average number $d$ of neighbors is then related to the covering radius and, consequently, to the transmitted power $P_{T}$, as
\begin{equation}\label{node_degree}
d=\pi r^2\varrho=\pi \frac{P_{T}}{P_{th}}\varrho
\end{equation}
where $\varrho$ is the spatial density of nodes inside a circle of area $\pi r^2$. In this setting, it is clear that the number $M$ of channels used to establish a communication must be designed with respect to the average number $d$ of neighbors, in order to keep the probability to have a collision among the communications of two nodes sufficiently small. Assuming independence among the channel selections of different nodes and exploiting (\ref{node_degree}), the probability that a packet is correctly exchanged over the selected channel is given by
\begin{equation}\label{P_c}
  p_c(M,P_T)=\left(\frac{M-1}{M}\right)^d=\left(1-\frac{1}{M}\right)^{\zeta P_T}
\end{equation}
where $\zeta=\pi\varrho/P_{th}$.  
As expected, the probability to establish correctly a communication link in (\ref{P_c}) gets worse by increasing the transmitted power $P_T$, because it translates in having more neighbors to communicate with, whereas, for a fixed transmitted power, it of course improves by taking a larger number of channels $M$.

\begin{figure}[t]
\centering
\includegraphics[width=8cm]{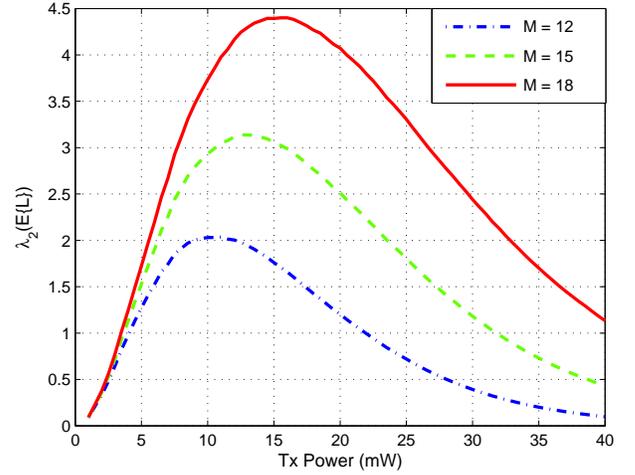}
  \caption{Behavior of $\lambda_2(\bar{\bL})$ versus the power transmitted by each node, for different number $M$ of available channels.}\label{mean_lambda_power}
\end{figure}
In an ideal communication case where no collisions occur, the increment of the power transmitted by each node leads to a monotonic increment of the network connectivity.
Thus, in an ideal case, it is always convenient to increase the power in order to increase the connectivity of the network, until full connectivity is reached. Nevertheless, in a real communication case, the presence of collisions due to the adoption of a random medium access protocol, e.g. the one introduced before, makes the graph describing the network a random graph, where each link is on with a probability given by (\ref{P_c}). It is then of interest to check the effect of collisions on the connectivity of the expected graph, which is actually the effective connectivity of the network. An example is given in Fig. \ref{mean_lambda_power}, where we show the behavior of the algebraic connectivity of the expected graph $\lambda_2(\bar{\bL})$ versus the power transmitted by each node, for different number $M$ of available channels. The simulation considers a network composed of $N=400$ nodes randomly deployed over a geographic area of $10^4$ m$^2$. The threshold power value at the receiver node is given by $P_{th}=0.01$ mW.
As we can notice from Fig. \ref{mean_lambda_power}, $\lambda_2(\bar{\bL})$ shows approximatively a quasi-concave behavior with respect to the transmitted power $P_T$. In fact, at low power values, the algebraic connectivity of the expected graph increases due to an increment of the links among neighbor nodes, whereas, at high power values, the number of neighbors becomes too large and the probability of having a collision increases, thus leading to a reduction of the overall connectivity of the network. From Fig. \ref{mean_lambda_power}, as expected, we also notice how, increasing the number of available channels $M$ for a fixed transmitted power, the connectivity of the expected graph improves. The behavior of $\lambda_2(\bar{\bL})$ shows that there is an optimal transmitted power that nodes should use to maximize the connectivity of the expected graph. An increment of the power with respect to this threshold value would lead to a waste of energy due to the effect of collisions, which becomes the dominant effect that drives to zero the connectivity. In summary, while in an ideal communication scenario nodes would always improve the network connectivity by increasing their transmitted power, considering a realistic random MAC, a too large transmission power may degrade the connectivity due to an increase in the collision probability.

\noindent{\it Distributed Connectivity Maximization:} In the previous section, we have shown that, for a sufficiently large number of nodes composing the network, the behavior of the algebraic connectivity of the expected graph $\lambda_2(\bar{\bL})$ versus the power $P_T$ transmitted by each node is unimodal, thus leading to the presence of a unique maximum point (see Fig. \ref{mean_lambda_power}). The goal of this section is to find the optimal power value $P_T^*$ that maximizes the connectivity of the expected graph, without assuming knowledge of the analytical relation between $\lambda_2(\bar{\bL})$ and $P_T$. The method is based on a stochastic algorithm that approximates the derivative of the function on the basis of noisy measurements of $\lambda_2(\bar{\bL})$.
For a given power value $P_T$, using the stochastic power iteration method in (\ref{lambdaN-1_2})-(\ref{lambda2_est}), it is possible to get an estimate $\hat{z}(P_T)$ of the second smallest eigenvalue of the expected Laplacian matrix in a totally distributed fashion. In practice, stopping the iterative method  in (\ref{lambdaN-1_2})-(\ref{pow_iter}) at a finite number of iterations, induces an inevitable estimation error, so that we can write
\begin{eqnarray}\label{noisy_measurements}
\hat{z}(P_T)=\lambda_{2}(\bar{\bL}(P_T))+\upsilon
\end{eqnarray}
where $\upsilon$ is a realization of a zero-mean random variable with bounded variance $\sigma^2_\upsilon$.
Now, exploiting the noisy measurements in (\ref{noisy_measurements}), we can use a Kiefer-Wolfowitz (KW) stochastic approximation  method \cite{Nevel} to find the maximum of the function $\lambda_{2}(\bar{\bL}(P_T))$. The algorithm  runs in parallel over  each node, which updates its own transmitted power according to the recursive rule:
\begin{eqnarray}\label{Kiefer_Wolfowitz}
P_T[t+1]=P_T[t]+q[t]\frac{\hat{z}(P_T[t]+c[t])-\hat{z}(P_T[t]-c[t])}{2c[t]}
\end{eqnarray}
$t\geq0$, where $P_T[0]$ is chosen at random, and $q[t]$ and $c[t]$ are two positive sequences that satisfy (\ref{Step_size}) and the further conditions
$c[t]\rightarrow 0$, and  $\sum_{t=0}^{\infty}\frac{q^2[t]}{c^2[t]}<\infty$.
For any $t$, the stochastic power iteration method in (\ref{lambdaN-1_2})-(\ref{lambda2_est}) must be run twice in order to get $\hat{z}(P_T[t]+c[t])$ and $\hat{z}(P_T[t]-c[t])$. The procedure in (\ref{Kiefer_Wolfowitz}) is then repeated until convergence.

\begin{figure}[t]
\centering
\includegraphics[width=8cm]{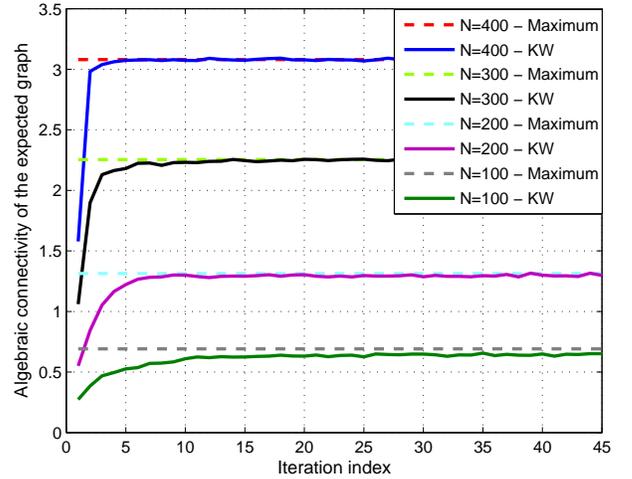}
  \caption{Behavior of $\lambda_2(\bar{\bL})$ versus iteration index, for different number $N$ of nodes.}\label{Max_conv}
\end{figure}
A numerical example is shown in Fig. \ref{Max_conv}, where we illustrate the behavior of the algebraic connectivity of the expected graph $\lambda_2(\bar{\bL})$ versus iteration index, for different values of the number of network nodes $N$, obtained by using the KW method in (\ref{Kiefer_Wolfowitz}). The maximum values of the algebraic connectivity are also reported for comparison purposes. The simulation considers a network composed of $N$ nodes randomly deployed over a geographic area of $10^4$ m$^2$. The threshold power value at the receiver node is given by $P_{th}=0.01$ mW. The number of channels is set to $M=15$. The step-size sequences are chosen as $q[t]=1/t$, and $c[t]=1/\sqrt[3]{t}$. As we can notice from Fig. \ref{Max_conv}, when the number of nodes is sufficiently large, the KW method in (\ref{Kiefer_Wolfowitz}) is able to find the maximum of $\lambda_{2}(\bar{\bL}(P_T))$ in a few iterations. At the same time, reducing the number of nodes in the network, the behavior of $\lambda_{2}(\bar{\bL}(P_T))$ looses its unimodality. This implies that the algorithm in (\ref{Kiefer_Wolfowitz}) can get stuck in some local maximum, thus explaining the gap between the maximum connectivity value and the KW method in Fig. \ref{Max_conv}, at low number of nodes $N$.

\section{Conclusions}

In this paper we have proposed a stochastic power iteration algorithm aimed at estimating the algebraic connectivity of an ad-hoc network in the case the communication among nodes is affected by random link failures.  We have proved that the algorithm converges almost surely to the second smallest eigenvalue of the expected Laplacian of the network graph and to its related eigenvector. Several numerical results confirm our theoretical findings. A distributed implementation of the proposed algorithm based on nested consensus loops is also proposed, and its performance is analyzed through several numerical simulations.
The proposed estimation method is then coupled with a power adaptation mechanism to implement a network connectivity
 control used to drive connectivity to a desired value. Finally, the behavior of the connectivity of the expected communication graph over a realistic MAC has been investigated. An interesting result is that, contrarily to what happens in the  absence of packet collisions, in a realistic scenario the nodes should not increase too much their transmission power, because this might heavily degrade the network connectivity due to an increase of collisions. Building on such a result, we have proposed a distributed KW stochastic approximation algorithm to find the transmit power that maximizes the connectivity in the presence of collisions.

\appendices

\section{Stochastic Approximation Theory}

In this section we report two theorems from stochastic approximation theory \cite{Nevel},\cite{Kar-Moura}, concerning the convergence properties of stochastic recursive procedures. We start introducing the notation from \cite{Nevel}, which is instrumental for the definition of the theorems. Let $\bw=\{\bw[k]\}$ be a Markov process on $\mathbb{R}^{N}$. Given  a nonnegative function
$V(k,\bw)$, the generating operator $\mathcal{L}V(k,\bw)$ is defined as
\begin{eqnarray}\label{Gen_operator}
\mathcal{L}V(k,\bw)=\mathbb{E}\left[ V(k+1,\bw[k+1]) | \bw[k]=\bw \right]-V(k,\bw)
\end{eqnarray}
$k\geq 0$, $\bw\in \mathbb{R}^{N}$, provided the conditional expectation exists. We say that $V(k,\bw)\in\mathcal{D}_\mathcal{L}$ in a domain $\mathcal{A}$, if $\mathcal{L}V(k,\bw)$ is finite for all $(k,\bw)\in \mathcal{A}$. Denoting the Euclidean distance between two points $\bx$ and $\by$ by $\rho(\bx,\by)$, the $\Delta$-neighborhood of a set $S$ and its complement are defined as
\begin{align}
  \begin{array}{ll}
    U_\Delta(S)&=\left\{\bw|\inf_{\mathbf{y}\in S} \rho(\bw,\by)<\Delta\right\},\vspace{.2cm} \\
    Q_\Delta(S)&=\mathbb{R}^{N} \backslash  U_\Delta(S).
  \end{array}
\label{neigh1}
\end{align}

We recall next two Theorems from \cite{Nevel} that will be useful in the sequel.

\noindent {\textit{Theorem 2 }:} Let $\bw$ be a Markov process with generating operator $\mathcal{L}$. Let there exist a nonnegative function $V(k,\bw)\in\mathcal{D}_\mathcal{L}$ in the domain $k\geq 0$, $\bw\in \mathbb{R}^{N}$, with the following properties:
\begin{align}
\inf_{k\geq0, \bw\in Q_\Delta(S)} \;&V(k,\bw)>0, \quad \forall \Delta>0 \\
&V(k,\boldsymbol{w})=0, \quad \bw\in S \\
\lim_{\boldsymbol{w}\rightarrow S} \sup_{k\geq 0}\; &V(k,\bw)=0  \\
\mathcal{L}V(k,\bw)&\leq g[k](1+V(k,\bw))-\beta[k]\phi(k,\bw) \label{LV}
\end{align}
where
\begin{align}
\inf_{k\geq0, \boldsymbol{w}\in Q_\Delta(S)}& \phi(k,\bw) >0, \;\forall \Delta>0 \\
&\hspace{-1cm}\beta[k]>0, \quad \sum_{k\geq 0} \beta[k]=\infty \\
&\hspace{-1cm} g[k]>0, \quad \sum_{k\geq 0} g[k]<\infty
\end{align}
Then, the Markov process $\bw=\{\bw[k]\}_{k\geq 0}$ with arbitrary initial distribution converges a.s. to $S$ as $k\rightarrow\infty$, i.e.
\begin{align}
\mathbb{P}\left(\lim_{k \rightarrow \infty} \rho(\bw[k],S)=0\right)=1,
\end{align}
with $\mathbb{P}(E)$ denoting the probability of the event $E$.

\begin{proof}
The proof can be found in \cite{Nevel},\cite{Kar-Moura}.
\end{proof}

\noindent {\textit{Theorem 3 }:}
Let $\{\bz[k]\}_{k\geq0}$ be a random vector generated by a Markov process defined by the difference equation
\begin{eqnarray}\label{Stoch_Recursion}
\bz[k+1]=\bz[k]+\alpha[k]\big[\bh(\bz[k])+\boldsymbol{\varphi}(k,\bz[k],\omega)\big]
\end{eqnarray}
with initial condition $\bz[0]=\bz_0$, where $\bh(\cdot):\mathbb{R}^{N}\rightarrow \mathbb{R}^{N}$ is Borel-measurable,
$\boldsymbol{\varphi}(k,\bz[k],\omega)$ is a family of zero-mean random vectors in $\mathbb{R}^{N}$, defined on some probability space $(\Omega,\mathcal{F},\mathcal{P})$, and $\omega\in\Omega$ is a canonical element of $\Omega$.
Consider the following set of conditions:

\noindent {\textit{Condition C.1} :} There exists a nonnegative function $V(\bz)\in C_2$ with bounded second-order partial derivatives and a point $\bz^*\in \mathbb{R}^{N}$ satisfying the conditions
\begin{align}
\displaystyle &V(\bz^*)=0, \quad V(\bz)>0\quad \hbox{for} \quad \bz\ne\bz^*\\
\displaystyle &\lim_{\|\boldsymbol{z}\|\rightarrow \infty}V(\bz)=\infty\\
\displaystyle &\sup_{\mu<\|\boldsymbol{z}-\boldsymbol{z}^*\|<1/\mu}(\bh(\bz),\mathbf{\nabla}_{\boldsymbol{z}}V(\bz))<0, \hspace{.3cm}\forall\hspace{.1cm} \mu>0, \label{Lyap_cond}
\end{align}
where $(\cdot,\cdot)$ denotes the inner product operator.\\
\noindent {\textit{Condition C.2} :} There exist two constants $k_1,k_2>0$ such that
\begin{align}
\|\bh(\bz)\|^2+\mathbb{E}\|\boldsymbol{\varphi}(k,\bz,\omega)\|^2\leq& \;\; k_1(1+V(\bz))\nonumber\\ &-k_2(\bh(\bz),\mathbf{\nabla}_{\boldsymbol{y}}V(\bz)) \label{Bound_variance}
\end{align}
\noindent {\textit{Condition C.3} :} The step-size sequence $\{\alpha[k]\}_{k\geq0}$ satisfies the persistence conditions in (\ref{Step_size}).

Let the conditions \noindent{\textit{C.1}}-\noindent{\textit{C.3}} hold for the process $\{\bz[k]\}_{k\geq0}$ given by (\ref{Stoch_Recursion}). Then, $\{\bz[k]\}_{k\geq0}$ is a Markov process and, starting from an arbitrary initial condition $\bz_0$, it converges almost surely to $\bz^*$ as $k\rightarrow \infty$.

\begin{proof}
The proof can be found in \cite{Nevel}.
\end{proof}

\section{Proof of Theorem 1}

The first part of the proof shows that the sequence $\bx[k]$ in (\ref{pow_iter}) converges a.s. to the normalized Fiedler vector $\hat{\bu}_2$. The power iteration in (\ref{pow_iter}) can be cast as:
\begin{equation}\label{pow_iter2}
   \displaystyle \bx[k+1]=\frac{\prod_{l=0}^k\bB_2[l]\bx_0}{\left\|\prod_{l=0}^k\bB_2[l]\bx_0\right\|}
\end{equation}
where $\bx_0$ is the initialization and $\bB_2[k]$ is given by (\ref{Defl_matrix3}).
Let us focus on the numerator of (\ref{pow_iter2}), which can be rewritten as:
\begin{align}\label{numer}
   \bw[k+1]=\prod_{l=0}^k\left(\bW_2[l]-\frac{1}{N}\mathbf{11}^T\right)\bx_0.
\end{align}
We first prove the boundness of the sequence $\bw[k]$ in (\ref{numer}). Exploiting the fact that $\bW_2[k]$ is a sequence of right stochastic matrices, we have
\begin{align}\label{numer2}
\bw[k+1]=\left(\bW_2[0]-\frac{1}{N}\mathbf{11}^T\right)\left(\prod_{l=1}^k\bW_2[l]\right)\bx_0.
\end{align}
Since $\bW_2[k]$ is a sequence of finite, primitive, (right) stochastic matrices, the sequence $\prod_{l=1}^k\bW_2[l]$ in (\ref{numer2}) has a finite limit as $k\rightarrow\infty$, \cite{Olfati1},\cite{Wolfowitz}. Thus, assuming a finite initial value $\bx_0$, from (\ref{numer2}) we have
\begin{align}\label{boundw}
   \|\bw[k]\|\leq c, \quad\quad \forall k,
\end{align}
where $c$ is a positive, finite constant.  Let us now introduce the notation ${\rm span}\{\bx\}=\{\hat{\bx}\in\mathbb{R}^{N\times 1}: \hat{\bx}=\zeta\bx,\zeta\in\mathbb{R}\}$. The proof follows by showing that the sequence $\bw[k]$ in (\ref{numer}) converges a.s. to ${\rm span}\{\bu_{2}(\bar{L})\}$, as $k\rightarrow\infty$, thus guaranteeing the a.s. convergence of the sequence $\bx[k]$ in (\ref{pow_iter2}) to the normalized Fiedler vector $\hat{\bu}_{2}(\bar{L})$, thanks to the presence of the normalization in (\ref{pow_iter2}). The sequence $\bw[k+1]$ in (\ref{numer}) can be equivalently recast as the following recursive rule:
\begin{align}\label{sequence2}
  \bw[k+1]=\left(\bI-\varepsilon[k]\bar{\bL}-\frac{1}{N}\mathbf{11}^T-\varepsilon[k]\tilde{\bL}[k]\right)\bw[k],
\end{align}
with $\bw[0]=\bx_0$. We will now use Theorem 2 to prove that the sequence in (\ref{sequence2}) converges to ${\rm span}\{\bu_{2}(\bar{L})\}$. To avoid overcrowding of formulas, in what follows we use the notation $\bu_i=\bu_i(\bar{\bL})$ and $\lambda_i=\lambda_i(\bar{\bL})$, $i=1,\ldots,N$, for the eigenparameters of the expected Laplacian, whereas the dependency is explicitly written in the case of other matrices. Under the assumption of temporal independence of the sequence $\tilde{\bL}[k]$, the sequence $\bw[k]$ in (\ref{sequence2}) is a Markov process. Let us define the potential function (independent of $k$):
\begin{align}\label{Potential_V2}
  V(\bw)=\bw^T\bF\bw,
\end{align}
where
\begin{align}\label{Matrix_F}
\bF:=\bar{\bL}+\frac{1}{N}\mathbf{11}^T-\lambda_2\bu_2\bu^T_2
\end{align}
is a positive semi-definite matrix such that its nullspace satisfies ${\rm Null}\{\bF\}={\rm span}\{\bu_2\}$. Let us then introduce the set $S={\rm span}\{\bu_2\}$. The potential function $V(\bw)\in\mathcal{D}_{\mathcal{L}}$ is non-negative. Since ${\rm Null}\{\bF\}=S$, we have
\begin{align}\label{cond1}
  V(\bw)=0, \; \bw\in S, \quad \lim_{\boldsymbol{w}\rightarrow S} V(\bw)=0.
\end{align}
The second condition in (\ref{cond1}) comes from the continuity of the potential $V(\bw)$. Let us consider the orthogonal decomposition $\bw=\bw_{S}+\bw_{S^\perp}$. Then, $\rho(\bw,S)=\|\bw_{S^\perp}\|$. By the definitions in (\ref{neigh1}), we have that $\bw\in Q_\Delta(S)$ implies $\|\bw_{S^\perp}\|\geq \Delta$. Hence, we obtain
\begin{align}
 \inf_{\boldsymbol{w}\in Q_\Delta(S)} V(\bw)\geq \lambda_2(\bF)\|\bw_{S^\perp}\|^2\geq \lambda_2(\bF)\Delta^2>0, \nonumber
\end{align}
with $\lambda_2(\bF)=\min(1,\lambda_3)$. Now, consider $\mathcal{L}V(\bx)$ in (\ref{Gen_operator}). Exploiting (\ref{sequence2}) and (\ref{Potential_V2}) in (\ref{Gen_operator}), and since $\mathbb{E}\{\tilde{\bL}[k]\}=\mathbf{0}$, we obtain
\begin{align}\label{Gen_operator2}
&\mathcal{L}V(\bw)=\mathbb{E}\left[ V(\bw[k+1]) | \bw[k]=\bw \right]-V(\bw) \nonumber\\
&=\bw^T\left(\bI-\varepsilon[k]\bar{\bL}-\frac{1}{N}\mathbf{11}^T\right)\bF\left(\bI-\varepsilon[k]\bar{\bL}-\frac{1}{N}\mathbf{11}^T\right)\bw \nonumber\\
& \quad+ \varepsilon^2[k]\bw^T\bP\bw -\bw^T\bF\bw
\end{align}
with $\bP=\mathbb{E}\left\{ \tilde{\bL}^T[k]\bF\tilde{\bL}[k] \right\}$. Let us now consider the following relations
\begin{align}
&\bF\mathbf{1}=\mathbf{1}, \quad \mathbf{1}^T\bF=\mathbf{1}^T, \label{eq_1}\\
&\bF\bar{\bL}=\bar{\bL}\bF=\bar{\bL}^2-\lambda^2_2\bu_2\bu_2^T, \\
&\bar{\bL}\bF\bar{\bL}=\bar{\bL}^3-\lambda^3_2\bu_2\bu_2^T, \\
&\bw^T\bP\bw\leq\lambda_{\max}(\bP)\|\bw\|^2\leq\eta\|\bw\|^2, \label{ineq_4}\\
&-\frac{1}{N}\mathbf{11}^T\leq -\varepsilon[k]\frac{1}{N}\mathbf{11}^T. \label{ineq_5}
\end{align}
The relation in (\ref{ineq_4}) holds true by the Gershgorin Theorem \cite{Horn-Johnson}, because the elements of the matrix $\tilde{\bL}[k]$ (thus, $\bP$) are taken from a finite set. Under assumption A.1, the sequence $\varepsilon[k]$ is positive and diminishing, and the inequality in (\ref{ineq_5}) holds true, for all $k$, if $\varepsilon[0]\leq1$. Otherwise, even if $\varepsilon[0]>1$, it exists a finite instant $k_0$ such that, for $k\geq k_0$, the inequality in (\ref{ineq_5}) is satisfied. Then, exploiting (\ref{boundw}) and (\ref{eq_1})-(\ref{ineq_5}) in (\ref{Gen_operator2}), we get
\begin{align}\label{Gen_operator3}
\hspace{-.15cm}\mathcal{L}V(\bw)\leq \varepsilon^2[k](\bw^T\bF_3\bw+\eta\cdot c)-2\varepsilon[k] \bw^T\bF_2\bw,
\end{align}
where
\begin{align}\label{F23}
\bF_2&=\bar{\bL}^2-\lambda^2_2\bu_2\bu^T_2+\frac{1}{2N}\mathbf{11}^T, \\
\bF_3&=\bar{\bL}^3-\lambda^3_2\bu_2\bu^T_2,
\end{align}
are positive semidefinite matrices. Considering the orthogonal decomposition $\bw=\bw_{S}+\bw_{S^\perp}$ and since ${\rm Null}\{\bF\}={\rm span}\{\bu_2\}=S$, we have
\begin{align}\label{ineqF3F}
\bw^T\bF_3\bw\leq \lambda^3_N\|\bw_{S^\perp}\|^2 \;\; \hbox{and} \;\;  \|\bw_{S^\perp}\|^2\leq\frac{\bw^T\bF\bw}{\lambda_2(\bF)}.
\end{align}
Thus, exploiting (\ref{ineqF3F}), the inequality in (\ref{Gen_operator3}) can be recast in the form of (\ref{LV}), where $\beta[k]=\varepsilon[k]$, and
\begin{align}
\phi(\bw)&=2\bw^T\bF_2\bw,\\
g[k]&=\varepsilon^2[k]\max\left(\frac{\lambda^3_N}{\lambda_2(\bF)},\eta\cdot c\right).
\end{align}
It is easy to see how the conditions of Theorem 2 on the sequences $\beta[k]$ and $g[k]$ are guaranteed by the choice of the sequence $\varepsilon[k]$ in (\ref{epsilonk}) made in Assumption A.1. Finally, since
$\bw\in Q_\Delta(S)$ implies $\|\bw_{S^\perp}\|\geq\Delta$, we obtain
\begin{align}
 \inf_{\boldsymbol{w}\in Q_\Delta(S)} \phi(\bw)\geq \lambda_2(\bF_2)\|\bw_{S^\perp}\|^2\geq \lambda_2(\bF_2)\Delta^2>0, \nonumber
\end{align}
with $\lambda_2(\bF_2)=\min(1/2,\lambda_3^2)$. All the conditions of Theorem 2 are then satisfied, thus guaranteing that the sequence $\bw[k]$ in (\ref{sequence2}) converges almost surely to $S={\rm span}\{\bu_2\}$ as $k\rightarrow\infty$. This result, combined with (\ref{pow_iter2}), ensures that
\begin{align}
\lim_{k\rightarrow\infty} \bx[k]=\hat{\bu}_{2},
\end{align}
almost surely (w.p.1), where $\hat{\bu}_{2}$ denotes the normalized Fiedler vector of the expected Laplacian matrix $\bar{\bL}$.
This completes the first part of the proof.

The second part of the proof aims to show the convergence of the sequence of estimates $z[k]$ in (\ref{lambda2_est}) to the algebraic connectivity $\lambda_{2}$. To prove it, we notice that, since $\lim_{k\rightarrow\infty}\bx[k]=\hat{\bu}_2$, and further we have $\|\hat{\bu}_2\|^2=1$, the behavior of (\ref{lambdaN-1_2}) at time $k$ can be written w.l.o.g. as
\begin{equation}\label{conv_y}
y_0[k]=\frac{\bx^T[k]\bB[k]\bx[k]}{\bx^T[k]\bx[k]}= \hat{\bu}_2\bB[k]\hat{\bu}_2+e[k]
\end{equation}
where $e[k]$, such that
\begin{equation}\label{eps_properties}
|e[k]|\leq e_0<\infty \;\; \forall k, \quad \hbox{and} \quad \lim_{k\rightarrow\infty}e[k]=0,
\end{equation}
is a random error due to the fact that, at time $k$, the power iteration in (\ref{pow_iter}) has not converged yet. Now, since $\displaystyle \bB[k]=\bar{\bW}-\frac{1}{N}\mathbf{1}\mathbf{1}^T-\bar{\varepsilon}\tilde{\bL}[k]$ and $\mathbf{1}^T\hat{\bu}_2=0$, from (\ref{conv_y}) we have
\begin{align}\label{conv_y2}
y_0[k]=\; &\hat{\bu}_2^T\bar{\bW}\hat{\bu}_2-\bar{\varepsilon}\cdot\hat{\bu}_2^T\tilde{\bL}[k]\hat{\bu}_2+e[k] \nonumber\\
=\; &\lambda_{N-1}(\bar{\bW})+e[k]-\bar{\varepsilon}\cdot\hat{\bu}_2^T\tilde{\bL}[k]\hat{\bu}_2
\end{align}
Then, letting $\tilde{n}[k]=-\hat{\bu}_2^T\tilde{\bL}[k]\hat{\bu}_2$, the a.s. behavior of the recursion in (\ref{lambdaN-1}) is given by
\begin{align}\label{asymptotic_recursion}
  \hspace{-.37cm}     y[k+1]=y[k]+\alpha[k]\big(\lambda_{N-1}(\bar{\bW})-y[k]+e[k]-\bar{\varepsilon}\tilde{n}[k]\big)
\end{align}
From (\ref{asymptotic_recursion}), exploiting expression (\ref{lambda2_est}), it is possible to derive the recursion for the sequence $z[k]$ as:
\begin{align}\label{asymptotic_recursion2}
 \hspace{-.2cm} z[k+1]= \hspace{.1cm}z[k]+&\alpha[k]\left(\lambda_{2}-z[k]+\tilde{n}[k]+e_{\varepsilon}[k]\right)
\end{align}
where $e_{\varepsilon}[k]=e[k]/\bar{\varepsilon}$. Since $e_{\varepsilon}[k]$ is a random quantity, it can be written w.l.o.g. as $e_{\varepsilon}[k]=\bar{e}_{\varepsilon}[k]+\tilde{e}_{\varepsilon}[k]$, where $\bar{e}_{\varepsilon}[k]$ is the mean part and $\tilde{e}_{\varepsilon}[k]$ is a zero-mean random fluctuation. Thus, (\ref{asymptotic_recursion2}) can be recast in the notation of Theorem 3, where:
\begin{align}
       h(z[k])&=\lambda_{2}-z[k]+\bar{e}_{\varepsilon}[k] \\
       \varphi(k,z[k],\omega)&=\tilde{n}[k]+\tilde{e}_{\varepsilon}[k] \label{Gamma_function}
\end{align}
The proof follows by showing that the process $\{z[k]\}_{k\geq0}$, generated by the recursion in (\ref{asymptotic_recursion}), satisfies the  conditions \textit{C.1}-\textit{C.3} of Theorem 3. Consider the filtration of the $\sigma$-algebra generated by the initial point $z[0]$ and the stochastic error sequence $\{\tilde{\bL}[l]\}$, for $0\leq l<k$, i.e.,
$\mathcal{F}_k=\sigma\left(z[0],\{\tilde{\bL}[l]\}_{0\leq l<k}\right)$.
The random family ${\varphi}(k,\cdot,\cdot)$ in (\ref{Gamma_function}) is $\mathcal{F}_{k+1}$ measurable, zero mean and independent of $\mathcal{F}_k$, thus making the random process $\{z[k],\mathcal{F}_k\}_{k\geq0}$ a Markov process. We will show now the existence of a potential function $V(z[k])$ such that the recursion in (\ref{asymptotic_recursion}) satisfies the conditions \textit{C.1}-\textit{C.3}. To this end, we define
\begin{eqnarray}\label{Potential_V}
V(z[k])=\left(\lambda_{2}-z[k]\right)^2.
\end{eqnarray}
It is easy to see how $V(z[k])\in C_2$ and is non-negative. Furthermore, setting $z^*=\lambda_{2}$, we have $V(z^*)=0$, $V(z[k])>0$ for $z[k]\ne z^*$, $\displaystyle \lim_{|z[k]|\rightarrow \infty}V(z[k])=\infty$, and
\begin{align}
&\sup_{\mu<|z[k]-z^*|<1/\mu}\left(h(z[k]),\frac{dV(z[k])}{dz}\right) \label{Lyap_cond}\\
&=\sup_{\mu<|z[k]-z^*|<1/\mu}-2(z^*-z[k])^2-2\bar{e}_{\varepsilon}[k](z^*-z[k]) \nonumber\\
&<-2\mu^2+\frac{2}{\mu}\bar{e}_{\varepsilon}[k]  \nonumber
\end{align}
Since $\displaystyle \lim_{k\rightarrow\infty}\bar{e}_{\varepsilon}[k]=\lim_{k\rightarrow\infty}\bar{e}[k]/\bar{\varepsilon}=0$ due to (\ref{eps_properties}), it exists a finite time instant $k_\mu$ such that, for $k\geq k_\mu$, the Lyapunov condition in (\ref{Lyap_cond}) is always satisfied, for any choice of the positive parameter $\mu$. To check condition \textit{C.2}, we note that
\begin{align}\label{Gamma}
\mathbb{E}|\varphi(k,&\omega)|^2\leq2\mathbb{E}\left|\tilde{n}[k]\right|^2+2\mathbb{E}\left|\tilde{e}_{\varepsilon}[k]\right|^2
\end{align}
From (\ref{eps_properties}), the error sequence $|\tilde{e}_{\varepsilon}[k]|^2=|\tilde{e}[k]|^2/\bar{\varepsilon}^2$ is upper bounded by a finite constant. Furthermore, the matrix $\tilde{\bL}[k]$ takes values from a finite set, thus implying that its eigenvalues are finite \cite{Horn-Johnson}. Thus, the variance in (\ref{Gamma}) can be upper bounded by a positive constant $c_2$, i.e., $\mathbb{E}|\varphi(k,\omega)|^2 \leq c_2$. Then, we have
\begin{align}
|h(z[k])|^2&+\mathbb{E}|\varphi(k,\omega)|^2\leq  (z^*-z[k])^2+\bar{e}^2_{\varepsilon}[k] \nonumber\\
&\hspace{3cm}+2\bar{e}_{\varepsilon}[k](z^*-z[k])+c_2 \nonumber\\
& \hspace{-1cm}=c_2+\bar{e}^2_{\varepsilon}[k] +\frac{1}{2}(z^*-z[k])^2+\frac{1}{2}(z^*-z[k])^2 \nonumber\\
& \hspace{1cm}+2\bar{e}_{\varepsilon}[k](z^*-z[k])  \nonumber\\
&\hspace{-1cm} \leq k_1(1+V(z[k]))-k_2\left(h(z[k]),\frac{dV(z[k])}{dz}\right)
\end{align}
where $k_1=\max\left(c_2+\frac{e^2_0}{\bar{\varepsilon}^2},\frac{1}{2}\right)>0$ and $k_2=1$.  This verifies also condition \textit{C.2} of Theorem 3 and condition \textit{C.3} is satisfied by the choice of $\{\alpha[k]\}_{k\geq0}$ in (\ref{Step_size}) made in Assumption A.1. The previous analysis shows that there exists a finite time instant $k_\mu$ such that, for $k\geq k_\mu$, all the conditions of Theorem 3 are satisfied, thus ensuring the convergence result
\begin{eqnarray}
\lim_{k\rightarrow\infty} z[k]= z^*=\lambda_{2} \quad \hbox{a.s.}
\end{eqnarray}
This concludes the proof of Theorem 1.

\section{Expected Convergence Rate}

Let us consider the behavior of $\mathbb{E}\bw[k]$ in (\ref{numer}):
\begin{equation}\label{Exp_numer}
   \mathbb{E}\bw[k+1]=\prod_{l=0}^k\bar{\bB}_2[l]\bx_0=\prod_{l=0}^k\left(\bar{\bW}_2[l]-\frac{1}{N}\mathbf{11}^T\right)\bx_0
\end{equation}
Now, since we have
\begin{equation}\label{eigen}
  \bar{\bW}_2[l]-\frac{1}{N}\mathbf{11}^T=\sigma_{N-1}[l]\bu_{N-1}\bu_{N-1}+\sum_{i=1}^{N-2}\sigma_i[l]\bu_i\bu_i^T \nonumber
\end{equation}
where $\sigma_{i}[l]=\lambda_{i}(\bar{\bW}_2[l])=1-\varepsilon[l]\lambda_{N-i+1}$ and $\bu_{i}=\bu_i(\bar{\bW}_2[l])=\bu_{N-i+1}$ is the corresponding eigenvector, the expression in (\ref{Exp_numer}) can be recast as
\begin{align}\label{Exp_numer2}
& \mathbb{E}\bw[k+1]=\left(\prod_{l=0}^k\sigma_{N-1}[l]\right)\cdot\bigg(\bu_{2}\bu_{2}^T \nonumber\\
&\quad\quad\quad+\sum_{i=1}^{N-2}\prod_{l=0}^k\left(\frac{\sigma_i[l]}{\sigma_{N-1}[l]}\right)\bu_{N-i+1}\bu_{N-i+1}^T\bigg)\bx_0
\end{align}
From (\ref{Exp_numer2}), neglecting the shrinking factor $\left(\prod_{l=0}^k\sigma_{N-1}[l]\right)$ that is canceled by the normalization of the power iteration, the slowest mode in (\ref{Exp_numer2}) is given by
\begin{align}\label{slow_mode}
c_r[k]=\prod_{l=0}^k\left(\frac{\sigma_{N-2}[l]}{\sigma_{N-1}[l]}\right)=\prod_{l=0}^k\left(\frac{1-\varepsilon[l]\lambda_3}{1-\varepsilon[l]\lambda_2}\right).
\end{align}
Since $(1-xa)/(1-xb)\leq \exp(-(a-b)x)$, for $x\geq0$ and $a>b$, the slowest mode $c_r[k]$ is bounded as:
\begin{align}\label{conv_mean_part}
c_r[k]&\leq\prod_{l=0}^k\exp\Big(-(\lambda_3-\lambda_2)\varepsilon[l]\Big) \nonumber\\
&=\exp \left(-(\lambda_3-\lambda_2)\sum_{l=0}^k\varepsilon[l]\right)
\end{align}

\section*{Acknowledgements}

The authors thank the anonymous reviewers for the detailed suggestions and corrections that improved the manuscript.

\end{document}